\documentclass[twocolumn,showpacs,preprintnumbers,amsmath,amssymb,prl]{revtex4-1}
\usepackage{epsfig} 
\usepackage{dcolumn}
\usepackage{bm}

\begin{document}

 
\title{Measurement of the lifetime of the $7s^2S_{1/2} $ state in atomic cesium using asynchronous gated detection}

\author{George Toh$^{1,2}$, Jose A. Jaramillo-Villegas$^{1,2,3}$, Nathan Glotzbach$^{1,4}$, Jonah Quirk$^{4,5}$, Ian C. Stevenson$^{1,2}$, J. Choi$^{1,2}$, Andrew M. Weiner$^{1,2}$, and D. S. Elliott$^{1,2,4}$}
\affiliation{%
   $^1$School of Electrical and Computer Engineering, Purdue University, West Lafayette, Indiana 47907, USA\\
   $^2$Purdue Quantum Center, Purdue University, West Lafayette, Indiana 47907, USA\\
   $^3$Facultad de Ingenier\'{i}as, Universidad Tecnol\'{o}gica de Pereira, Pereira, Risaralda 660003, Colombia\\
   $^4$Department of Physics and Astronomy, Purdue University, West Lafayette, Indiana 47907, USA\\ 
   $^5$Pott College of Science, Engineering and Education, University of Southern Indiana, Evansville, Indiana 47712, USA 
}

\date{\today}

\begin{abstract}
We report a measurement of the lifetime of the cesium $7s\,^2S_{1/2} $ state using time-correlated single-photon counting spectroscopy in a vapor cell. We excite the atoms using a Doppler-free two-photon transition from the $6s\,^2S_{1/2}$ ground state, and detect the 1.47~$\mu$m photons from the spontaneous decay of the $7s\,^2S_{1/2}$ to the $6p\,^2P_{3/2}$ state. We use a gated single photon detector in an asynchronous mode, allowing us to capture the fluorescence profile for a window much larger than the detector gate length. Analysis of the exponential decay of the photon count yields a $7s\,^2S_{1/2} $ lifetime of 48.28~$\pm$~0.07~ns, an uncertainty of 0.14\%. These measurements provide sensitive tests of theoretical models of the Cs atom, which play a central role in parity violation measurements.
\end{abstract}

                             
 \maketitle

Precision laboratory measurements of electric dipole (E1) matrix elements are critical for the advancement of atomic parity violation (PV) studies in several regards:  Precise models of atomic structure are required to extract the weak charge $Q_w$ from any measurement of the PV transition moment; E1 matrix elements are included explicitly in the perturbative expansion for the PV moment; and measurements of the PV amplitude are always carried out relative to a different optical transition amplitude, such as a Stark-induced amplitude.  Thus, we require precise determinations of electric dipole matrix elements, through a variety of laboratory measurements, and detailed comparison with {\it ab initio} theoretical results.  

The most precise determination of a PV moment in any atomic system is that of the $6s\,^2S_{1/2} \rightarrow 7s\,^2S_{1/2} $ transition in cesium, carried out by Wood {\it et al.~}in 1997~\cite{WoodBCMRTW97}.
In the past 30 years, several advances in models of the atomic structure of the cesium atom~\cite{DzubaFS89,DzubaFKS89,BlundellSJ92, DzubaFS97,SafronovaJD99,Derevianko00, DzubaF00,DzubaFG02,PorsevBD09,PorsevBD10,DzubaBFR12}, 
and measurements of key transition amplitudes~\cite{BouchiatGP84,TannerLRSKBYK92, YoungHSPTWL94,Hoeling:96,DiBerardinoTS98,RafacT98,RafacTLB99,VasilyevSSB02,AminiG03,ZhangMWWXJ13, antypas7p2013, antypasm12013,toh7p2014,PattersonSEGBSK15,GregoireHHTC15} have been reported.  The uncertainty in the E1 transition moment $\langle 7s||r||6p_{1/2} \rangle$ is presently one of the primary contributors, along with the $\langle 7p_{1/2}||r||6s \rangle$ matrix element, to the uncertainty in the PV moment for the $6s\,^2S_{1/2} \rightarrow 7s\,^2S_{1/2} $ transition~\cite{PorsevBD10,antypas7p2013}.  Similarly, the uncertainties in $\langle 7s||r||6p_{1/2} \rangle$ and $\langle 7s||r||6p_{3/2} \rangle$ are primary contributors to the uncertainty of the scalar Stark polarizability for the $6s \rightarrow 7s $ transition~\cite{VasilyevSSB02,antypas7p2013}.  

In this paper we present our measurement of the lifetime of the cesium $7s\,^2S_{1/2} $ state using an asynchronous time-correlated single-photon counting (TCSPC) technique.  
By measuring the lifetime of the 7s state, we indirectly measure the matrix elements named above.
We find a lifetime value of 48.28~$\pm$~0.07~ns, in good agreement with the previous measurement by Bouchiat {\it et al.}~\cite{BouchiatGP84}, but with much smaller uncertainty, and in agreement with several theoretical determinations~\cite{DzubaFKS89,BlundellSJ92,DzubaFS97,SafronovaJD99,DzubaFG02,PorsevBD10}. 
This work paves the way to reducing the uncertainty of the PV transition amplitude and Stark polarizability, and complements progress we are making toward a new atomic PV measurement in cesium \cite{choipnc2016,antypasm12013}.  

\begin{figure} [b!]
	  \includegraphics[width=8cm]{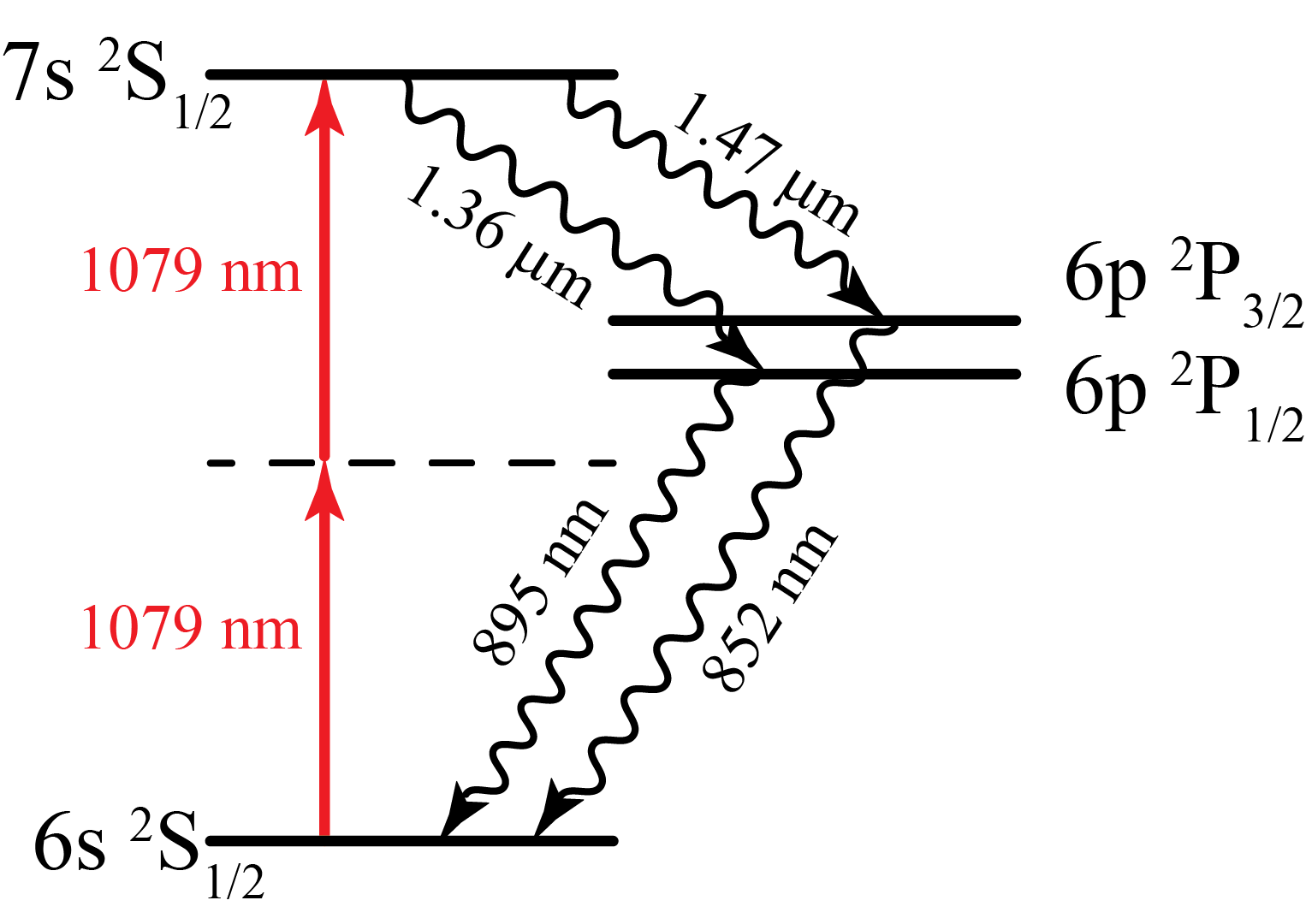}\\
	  \caption{Energy level diagram of atomic cesium, showing the states relevant to this experiment. Atoms are excited from the $6s\, ^2S_{1/2}$ ground state to the $7s\, ^2S_{1/2}$ excited state by two-photon excitation. Fluorescence photons at 1.47~$\mu$m from the decay of atoms from the 7s state to the $6p\,^2P_{3/2}$ state are collected and counted by the single photon detector. }
	  \label{fig:EnergyLevel}
\end{figure}

Cesium atoms in the $7s\, ^2S_{1/2}$ state can spontaneously decay through the $6p\, ^2P_{1/2}$ or $6p\, ^2P_{3/2}$ states, which subsequently decay to the $6s\, ^2S_{1/2}$ ground state, as shown in Fig.~\ref{fig:EnergyLevel}.  
The total decay rate $1/\tau_{7s}$ of the excited state is written as the sum of transition rates to these two intermediate states
  \begin{equation}
     \frac { 1 }{ { \tau_{7s}  } } = \sum_{J = 1/2, 3/2}     \frac { 4 }{ 3 }  \frac { { \omega_{J}}^{ 3 } }{ { c }^{ 2 } } \alpha \frac { { \left| \langle 7s||r||6p_{J} \rangle \right|  }^{ 2 } }{ 2J^{\prime}+1 } ,
   \end{equation}
where $\tau_{7s}$ is the lifetime of the $7s\,^2S_{1/2}$ state, $\omega_{1/2}$ and $\omega_{3/2}$ are the transition frequencies of the $7s\,^2S_{1/2} \rightarrow 6p\,^2P_{1/2}$ and $7s\,^2S_{1/2} \rightarrow 6p\,^2P_{3/2}$ transitions, respectively, $J^{\prime} = \frac{1}{2}$ is the angular momentum of the $7s\,^2S_{1/2}$ upper state, $\alpha$ is the fine-structure constant, and $c$ is the speed of light.  Once the lifetime of the $7s$ state is measured, only the ratio of matrix elements, $\langle 7s||r||6p_{3/2} \rangle / \langle 7s||r||6p_{1/2} \rangle$, is needed to extract the individual matrix elements.  This ratio is reliably calculated by theory and very consistent across different theoretical calculations~\cite{DzubaFKS89,BlundellSJ92,DzubaFS97,SafronovaJD99}.  

\begin{figure}
      \includegraphics[width=8cm]{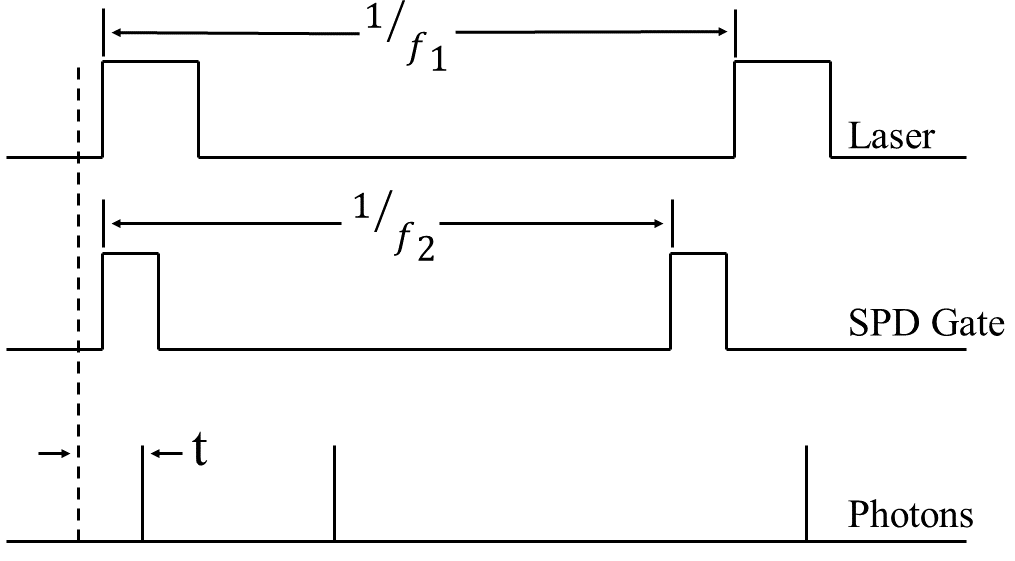} 
	  \caption{Timing diagram of the experiment.  
The dashed line represents the start time for the TCSPC module, and $t$ the arrival time of the first photon detected within the gate pulse.  
$f_1 = 1.25$ MHz is the laser repetition rate and $f_2$ is the SPD gate repetition rate.   
The difference in frequencies $(f_2 \neq f_1)$ causes the SPD to gate during a different part of the measurement window every cycle.  This gate-free method of capturing data allows us to utilize the SPD with a 40 ns gate, while capturing a 800 ns measurement window of photon fluorescence.}
	  \label{fig:TimingDiagram}
\end{figure}

\begin{figure} [b]
	  \includegraphics[width=8cm]{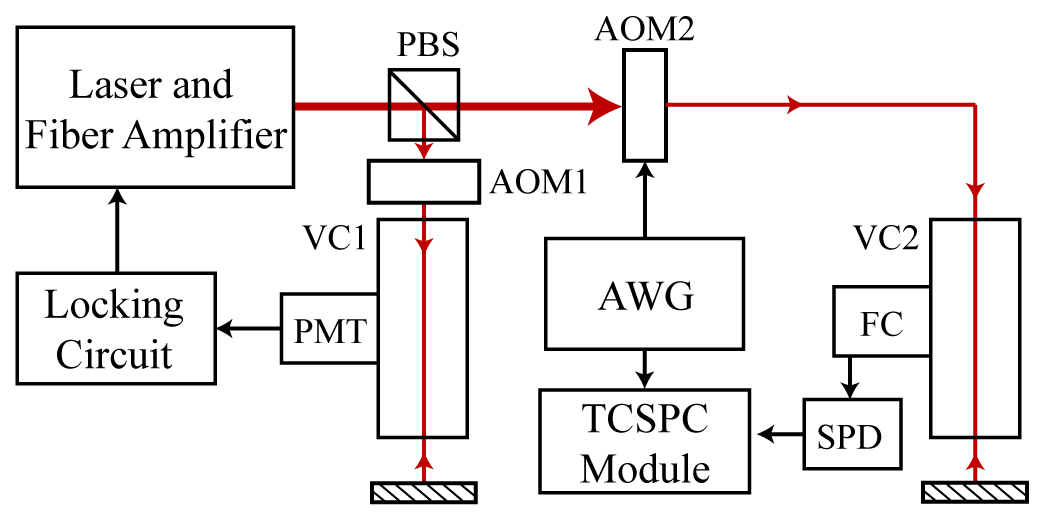}\\
	  \caption{Experimental setup.  Abbreviations in this figure are: (PBS) polarizing beam splitter cube; (AOM1) and (AOM2) acousto-optic modulators; (VC1) and (VC2) cesium vapor cells; (PMT) photomultiplier; (FC) fiber coupling optics; (AWG) arbitrary waveform generator; (SPD) single photon detector; and (TCSPC) time-correlated single photon counter. }
	  \label{fig:ExperimentalSetup}
\end{figure}

TCSPC has been used to accurately measure atomic excited state lifetimes in Cs~\citep{Hoeling:96,YoungHSPTWL94,DiBerardinoTS98}, Fr~\citep{zhao1997,simsarian1998,gomezfr2005} and Rb~\citep{sheng2008,gomezrb2005}.  A train of laser pulses repeatedly excites the atoms, and a detector records the exponential decay of fluorescence photons from the excited atoms. We introduce an asynchronous detection scheme in order to collect the fluorescence for a measurement window much longer than the gate duration of our gated single photon detector (SPD), and to reduce the impact of any possible temporal variations of the detector efficiency over the measurement window. 
The key to the asynchronous detection scheme is to cycle the laser excitation pulses and gated-SPD at different frequencies, $f_1$ and $f_2$, respectively, as illustrated in Fig.~\ref{fig:TimingDiagram}.  This causes varying delay times between the beginning of the measurement window (of duration $1/f_1$) and the SPD gate, which effectively causes the SPD gate pulse to repetitively scan across the full measurement window.  When repeated over many cycles, the result is a flat response of the detector in time, comparable to using a free-running detector~\cite{tosi2013}. 

We show a schematic of our experimental setup in Fig.~\ref{fig:ExperimentalSetup}. 
The excitation laser is a home-made 1079 nm external cavity diode laser (ECDL), coupled into a fiber amplifier to amplify the optical power to 4 W, and split along two paths using a polarizing beam splitter (PBS) cube. We use the first of these beams to lock the laser frequency to the two-photon resonance frequency, and the second to carry out the lifetime measurements.  The first beam passes through an acousto-optic modulator (AOM) driven by a constant-amplitude 90 MHz signal.  We direct the first-order diffracted beam to a heated vapor cell (VC1), where a photomultiplier tube (PMT) picks up atomic fluorescence at 852 nm.  This signal is processed and fed back to the laser frequency control to stabilize the laser frequency to the cesium $6s\,^2S_{1/2}, \ F=4 \rightarrow 7s\,^2S_{1/2}, \ F=4$ transition ($F$ is the total angular momentum, electron spin plus nuclear spin).
We direct the second beam from the PBS to a second AOM, which is also driven at 90 MHz. 
The rf power driving AOM2 is pulsed on for 250 ns at a repetition rate of $f_1 = 1.25$ MHz. 
This pulsed beam is focused into a second heated cesium vapor cell (VC2) in a nearly-counter-propagating geometry for Doppler-free two-photon excitation (for enhancement of the signal) of the $7s\,^2S_{1/2}$ state

We filter the fluorescence at 1.47~$\mu$m from this cell using a long-pass filter to reduce unwanted background (scattered laser light, other fluorescence components, and room lights, for example), and use a commercial fiber collimator to couple the fluorescence light into a $10~\mu$m single-mode fiber. We choose to detect this fluorescence line for its reduced susceptibility to radiation trapping effects, its time dependence as a simple single exponential (in contrast to the double exponential of~\cite{Hoeling:96,gomezfr2005,sheng2008,gomezrb2005}) and its large branching ratio, compared to the 1.36~$\mu$m line. 
The collection optics allows us to image decaying atoms within an area of $\sim500 \ \mu$m diameter. This detection volume is much greater than the region excited by the laser, and much larger than the $\sim$~10~$\mu$m distance traveled by an average velocity atom within one lifetime $\tau_{7s}$. 
The fiber transmits the fluorescence light to an Aurea Technology InGaAs gated avalanche single photon detector.

For accurate timing of photon arrivals, we use a HydraHarp 400 TCSPC module with a specified timing uncertainty of $<$12 ps.  An arbitrary waveform generator (AWG) produces the start pulse for the TCSPC module, indicating the start of the $1/f_1 =$ 800 ns long measurement window.  The AWG also generates the 90 MHz rf modulation pulse for driving AOM2, which generates the train of optical excitation pulses sent to VC2.  We gate the SPD on for $T_{\rm gate} =$ 40 ns at a slightly different frequency $f_2$ (where $f_2 \approx f_1 + 20 $ Hz).  
The TCSPC module registers the arrival time $t$ of a SPD pulse generated by the 1.47 $\mu$m fluorescence photon arriving within a gate pulse.
The precision of the lifetime measurement relies on the accuracy of the TCSPC timing module, but not on that of the frequency sources. 

\begin{figure}[b!]
      \includegraphics[width=\columnwidth]{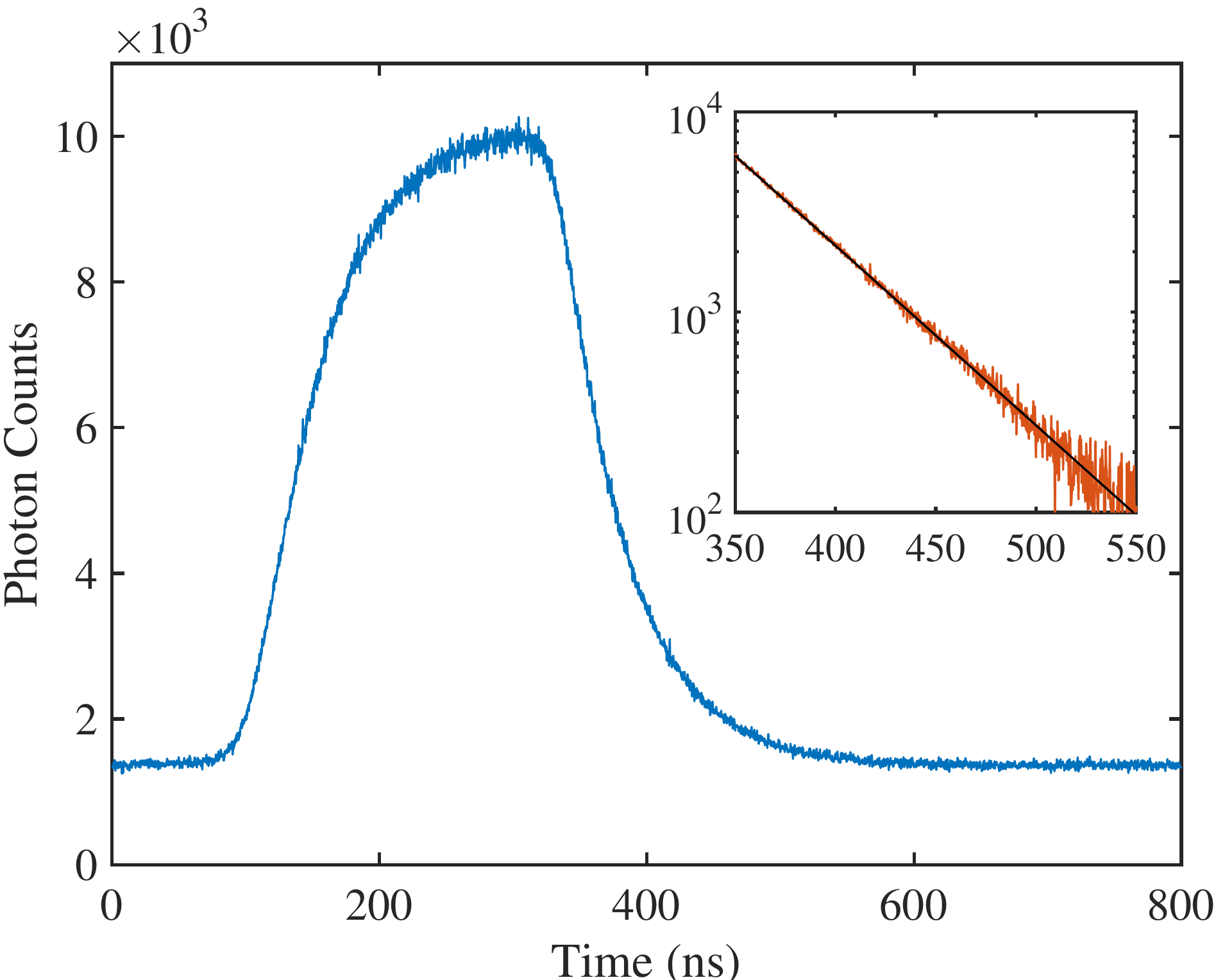}
	  \caption{Decay curve of the Cs $7s$ level. The main figure consists of 1 hour of recorded data and shows the excitation of atoms and exponential decay of fluorescence. \textbf{Inset:} The same data for $350 - 550$ ns with the background deducted, shown on a logarithmic scale in red and the best-fit line in black.}
	  \label{fig:LinLogFit}
\end{figure}
    
We show an example of the histogram of photon counts vs.\ $t$ in Fig.~\ref{fig:LinLogFit}.  In this figure, the ordinate represents the number of fluorescence photons $N_i$ detected in the \textit{i}-th bin over the course of a 1 hour data run, where each bin is of duration $T_{\rm bin} =$ 256 ps.  The laser turns on at $t \sim 100$ ns in this plot, and the fluorescence count approaches a steady-state value of $\sim 10,000$ counts per bin over the course of a few excited state lifetimes.  This corresponds to a photon incidence rate (without gating) of $2 \times 10^5$ per second, or the probability of detecting a photon within a 40 ns window of 0.8$\%$.  The laser then turns off at $\sim$320 ns, and the signal drops, approaching a baseline value which primarily represents the detector dark noise counts.  The noise level of our signal is consistent with the shot noise limit.

We apply two corrections to the raw data before determining the lifetime $\tau_{7s}$. The first is for pile-up error, in which we account for the probability that a second photon arrives within the 40 ns gated detection window.
The correction that we apply in the asynchronous measurement scheme differs from the typical pile-up error corrections described, for example, in \cite{gomezrb2005,simsarian1998,sheng2008}.
The probability of detecting a photon within the 40 ns window centered on the \textit{i}-th bin of the data set is approximately:
\begin{equation}
P_i = \frac{N_i}{N_E} \times \frac{T_{\rm gate}}{T_{\rm bin}} \times \left( \frac{1}{T_{\rm gate} f_1 } \right) = \frac{N_i}{N_E T_{\rm bin} f_1} ,
	\label{eqn:photonprobability}
\end{equation}
where $N_E$ is the total number of laser pulse repetitions (typically $f_1 \times 1 \ {\rm hour} = 4.5 \times 10^9$), and $T_{\rm gate} f_1$ is the duty cycle of the SPD gate. 
We make sure that $P_i < 1\%$ during peak fluorescence (when the laser pulse is on) to keep any needed corrections small.  For any gate pulse in which we detect a fluorescence photon, the probability of there being a second photon within that window is $P_i/2$.  This second photon is not detected, so we must multiply each point within the data set by $1 + P_i/2$.

We must also apply a correction to the data to account for the detector dead time. 
Because the detector dead time (1 $\mu$s) is longer than the timing window (0.8 $\mu$s), after a photon is detected, the gated-SPD is not ready to detect any photons during the next laser pulse cycle.  
We chose the frequency $f_1$ as a compromise between rapid data collection rates and long duration measurement windows, $1 / f_1 \gg \tau_{7s} \approx$ 50 ns. 
This necessitates an additional correction to the raw data of $1 + P_i$.  
In total, these two corrections alter the fitted lifetime by $0.2\%$.  

We fit an exponential function of the form
\begin{equation}
N_i ={ A }_{ 7s }\exp\left( -\frac { t  }{ { \tau  }_{ 7s } }  \right) + y_o  
\label{eq:fitfunction}
\end{equation}
to the falling edge of the data to extract the lifetime of the 7s state, $\tau_{7s}$. Here, $A_{7s}$ is the amplitude of the exponential and $y_0$ is the background photon count.
We show an example of data and the fitted function on a semi-log plot in the inset of Fig.~\ref{fig:LinLogFit}. The laser pulse has finite turn-off time, which we measured to be $\sim$20 ns ($90\% \: \textrm{to} \: 10\%$). This produces some ambiguity regarding the appropriate range of data to include in the fits, as the fluorescence decay follows an exponential only when the laser has completely turned off.  
We run fits to the data for a range of starting truncation points $t = 360 - 380$ ns, but use a fixed ending truncation point at $t =$ 800 ns.  
For each individual dataset, we determine the lifetime from the mean of these fitted lifetimes. The statistical uncertainties of these fits do not vary much across this 20 ns range, so we use the statistical uncertainty of the middle value, which we add in quadrature to the standard deviation over this range of lifetimes (the truncation error) to determine the uncertainty for each dataset. This effectively adds truncation error into our statistical uncertainty value. For most of the data sets, the truncation error is $\sim 50 \%$ of the statistical uncertainty.

\begin{figure}
      \includegraphics[width=\columnwidth]{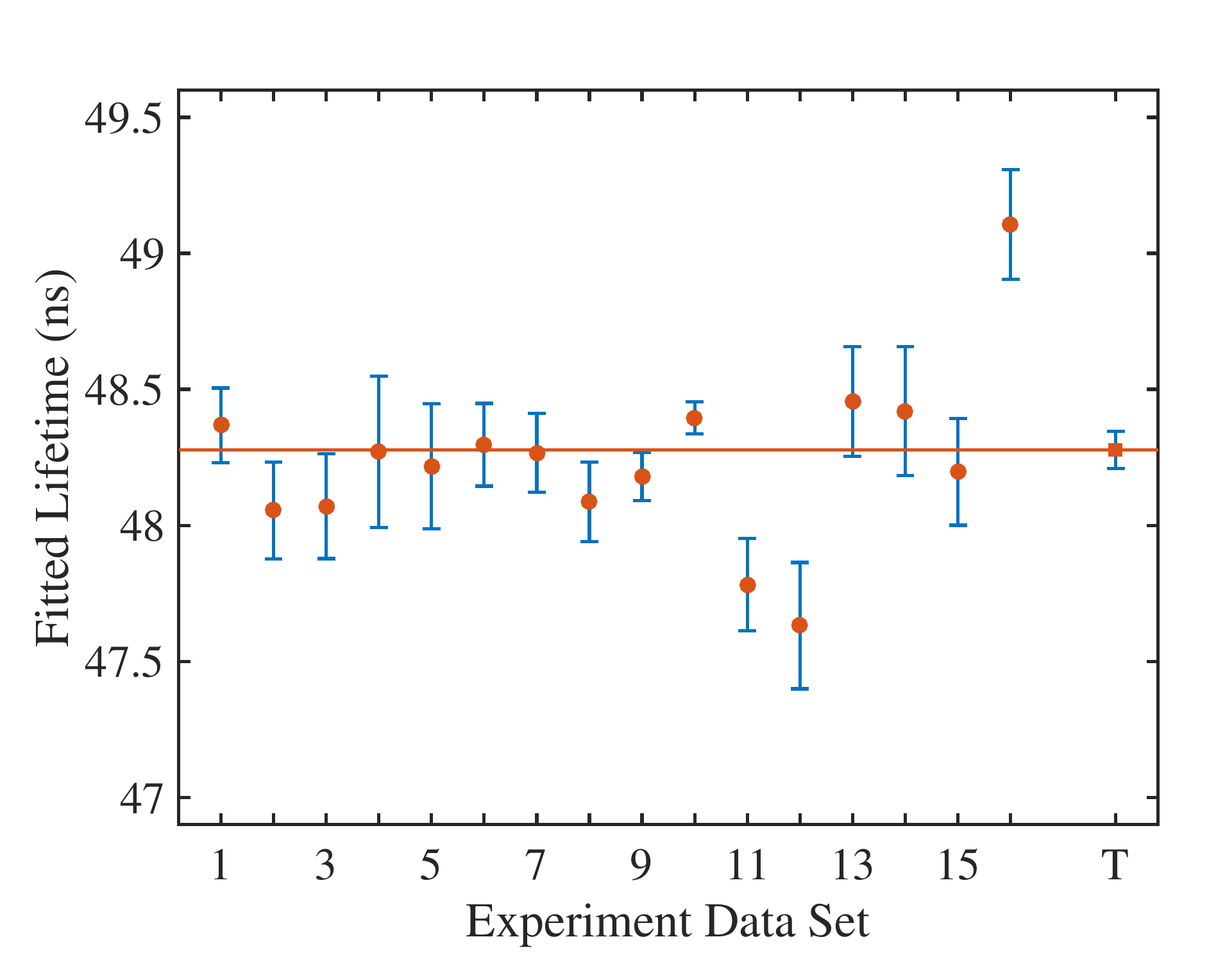}\\
	  \caption{A plot showing the 16 individual measurement results used to calculate the final value. Data sets 9 and 10 were 10 hours long, while the rest were for 1 hour. The total of 34 hours of data was captured over a period of three days. The final data point T and the red horizontal line is the weighted mean of the 16 data sets, with error bars inclusive of truncation and systematic uncertainties.} 
	  \label{fig:fittedresult}
\end{figure}

We show a plot of the 16 different measurement results used to calculate the final value of the $7s$ lifetime in Fig.~\ref{fig:fittedresult}. Fourteen 1 hour long data sets and two overnight data sets of 10 hours (labeled 9 and 10 in Fig.~\ref{fig:fittedresult}) were used to determine the final lifetime. 
The weighted mean of these 16 lifetimes is 48.28~$\pm$~0.03~ns. The reduced $\chi^2_\nu$ of the resulting fit was 2.98, suggesting that our uncertainties were not sufficiently conservative. We observed that the laser lost lock several times during runs $11-16$, which could be the cause of the larger variability of the results. For lack of a clear link however, we chose to increase our statistical uncertainty by $\sqrt{2.98}$.

In order to make a measurement with high accuracy, we investigated several potential systematic effects to determine their impacts on the measurement.  We verified that our measurement scheme counts photons at all times with equal probability (i.e.~there is no temporal variation in the detection sensitivity) by recording the background photon counts with the laser off. We measured the lifetime at several different cell temperatures and with different applied magnetic fields to verify that there was no effect from radiation trapping, collisions, or Zeeman quantum beats. (Data sets 6 through 9 of Fig. \ref{fig:fittedresult} were taken at a temperature of $\sim 127^{\circ}$C, with the rest taken at $\sim 118^{\circ}$C. In data sets 3 through 5, a 3 G magnetic field was applied to the vapor cell in each of three orthogonal directions.) Additionally, we quantified the effect of the detector jitter, included a correction for pile-up error and addressed truncation effects.  We summarize the magnitudes of these effects on our error budget in Table~\ref{table:sourcesoferror}. 
Adding statistical and systematic errors in quadrature, our final result is $\tau_{7s}=$ 48.28~$\pm$~0.07~ns. We display this final result as the last point in Fig.~\ref{fig:fittedresult}.

\begin{table}
    \begin{tabular}{l c}
        \hline\hline
        \multicolumn{1}{c}{Error}      & \% uncertainty   \\ \hline
        Statistical	and truncation     & 0.12             \\
        Detection sensitivity	       & 0.05             \\
        Radiation trapping			   & 0.03             \\
        Time calibration               & 0.03             \\
		Pile-up correction             & 0.02             \\
		SPD detector jitter			   & 0.01             \\
                                       &  		  		  \\
        Total uncertainty			   & 0.14             \\
    \hline \hline 
    \end{tabular}
    \caption{Sources of error and the percentage uncertainty resulting from each source. The error is dominated primarily by statistical error.}
    \label{table:sourcesoferror}
\end{table}

\begin{table}[t]
  \begin{tabular}{|l|c|}
    \hline
      \multicolumn{1}{|c|} {Group}   		& \hspace{0.15in}$\tau_{7s}$ (ns) \hspace{0.15in} \\ \hline 
  {\underline{\emph{Experimental}}} 		&         \\
   Marek, time-resolved fluorescence, 1977~\cite{Marek77}  & $  49  \pm 4 $  \\
   Hoffnagle {\it et al.}, Hanle effect, 1981~\cite{HoffnagleTW81}  & $  53.6  \pm 1.2 $  \\
   M. Bouchiat {\it et al.}, Hanle effect, 1984~\cite{BouchiatGP84}  & $  48.5  \pm 0.5 $  \\
   This work, time-resolved fluorescence                              & $  48.28  \: \pm 0.07 $  \\
   & \\
  \multicolumn{1}{|l|}{\underline{\emph{Theoretical}}} & \\
    C. Bouchiat {\it et al.}, semi-empirical~\cite{BouchiatPP83}							&  48.35   \\
    Dzuba {\it et al.},$^{\ast}$ 1989~\cite{DzubaFKS89}	&  48.07  \\
    Blundell {\it et al.},$^{\ast}$ 1992~\cite{BlundellSJ92}  &  48.56  \\
    Dzuba	{\it et al.},$^{\ast}$ 1997~\cite{DzubaFS97}	 &  48.07  \\
    Safronova {\it et al.},$^{\ast}$ 1999~\cite{SafronovaJD99}&  48.42  \\  
    Dzuba {\it et al.},$^{\dagger}$ 2002~\cite{DzubaFG02}	&  48.24  \\
    Porsev {\it et al.},$^{\dagger}$ 2010~\cite{PorsevBD10} &  48.33  \\
   \hline
  \end{tabular}
  \caption{Experimental and theoretical results for the lifetime $\tau_{7s}$ of the cesium $7s\,^2S_{1/2}$ state.  We derived theory values marked with an asterisk ($^{\ast}$) from matrix elements $\langle 7s||r||6p_{1/2} \rangle$ and $\langle 7s||r||6p_{3/2} \rangle$ reported here.  In the theoretical works marked with a dagger ($^{\dagger}$), the authors only reported values of $\langle 7s||r||6p_{1/2} \rangle$, so we estimated $\langle 7s||r||6p_{3/2} \rangle$ from $1.528 \times \langle 7s||r||6p_{1/2} \rangle$ in order to derive $\tau_{7s}$.}
  \label{table:ResultComparison}
\end{table}

We present a summary of past theoretical and experimental results in Table~\ref{table:ResultComparison}.
Our final result agrees well with the last experimental result by Bouchiat {\it et al.} \citep{BouchiatGP84} which was based on the Hanle effect.
The theory values shown in the table are calculated from the E1 matrix elements reported in these works and the measured transition energies.  Our result agrees within our uncertainty with the two most recent theoretical works by Dzuba \citep{DzubaFG02} and Porsev \citep{PorsevBD10}.  These works only report values of $\langle 7s||r||6p_{1/2} \rangle$, so we estimate the ratio $\langle 7s||r||6p_{3/2} \rangle / \langle 7s||r||6p_{1/2} \rangle = 1.528$ from the earlier theory papers~\cite{DzubaFKS89,BlundellSJ92,DzubaFS97,SafronovaJD99} to derive the lifetimes listed.  

In summary, we present a new lifetime measurement technique using a gated SPD in an asynchronous measurement scheme, and a new, higher precision measurement result for the lifetime of the $7s\,^2S_{1/2}$ state of cesium. This measurement technique allows us to collect data for a time window much longer than the maximum gate length of a gated SPD with uniform detection sensitivity. The scheme presented here can be used to measure atomic lifetimes with high precision. 
Our newly measured value of this lifetime agrees well with earlier experimental and theoretical determinations of the Cs $7s \, ^2S_{1/2}$ lifetime, and improves on the experimental uncertainty by a factor of seven.  The lifetime measurement result presented here tests models of the cesium atomic structure, and can be used to reduce uncertainties on the PV moment and the scalar polarizability for the $6s\,^2S_{1/2} \rightarrow 7s\,^2S_{1/2}$ transition.

This material is based upon work supported by the National Science Foundation under Grant Number PHY-1607603 and PHY-1460899. JAJ acknowledges support by Colciencias Colombia through the Francisco Jose de Caldas Conv.~529 Scholarship and Fulbright Colombia. We gratefully acknowledge useful discussions with M. Y. Shalaginov, M. S. Safronova and D. E. Leaird. 

\bibliography{biblio}

\end{document}